\documentclass[review]{elsarticle}

\usepackage{lineno,hyperref,amsmath,amssymb,bm}

\journal{Results in Physics}

\begin{document}
	
	\begin{frontmatter}
		
		\title{A novel idea: calculating anisotropic turbulence only by Kolmogorov structure constant $C_n^2$ and power law $\alpha$}

		\author{Chao Zhai\corref{cor1}}
		\cortext[cor1]{Corresponding author.}
		\ead{zhaichao@qhd.neu.edu.cn}
		
		\address{School of Computer and Communication Engineering, Northeastern University at Qinhuangdao, Qinhuangdao 066004, China}

		\begin{abstract}
			As stated in the present research, the anisotropic non-Kolmogorov turbulence model is more in line with the actual turbulence. However, anisotropic factor is vital for the accuracy of calculations, and measuring it precisely is a challenging task. In this paper, utilizing three modeling approaches, we derive four different equivalence formulas between the anisotropic non-Kolmogorov and the Kolmogorov structure constants, and apply them to the extant anisotropic results. It is found that the anisotropic factor is eliminated and the anisotropic turbulence can be predicted only by Kolmogorov structure constant and power law.
		\end{abstract}
		
		
	\end{frontmatter}

%%%%%%%%%%%%%%%%%%%%%%%%%%  body  %%%%%%%%%%%%%%%%%%%%%%%%%%
\section{Introduction}

It is widely recognized that atmospheric turbulence severely degrades the performance of laser absorption spectroscopy, laser imaging, and free-space optical communications \cite{Ricklin2006,Kazaura2006,Guo2019,Li2018}. In the past few decades, the Kolmogorov turbulence model was extensively utilized in the study of light waves transmission in turbulent atmosphere \cite{Tatarskii1961,Andrews2005}. However, recently due to the progress of measurement methods, both the experimental data and the theoretical results have exhibited that the anisotropic non-Kolmogorov turbulence model can better characterize the eddy impacts on light waves transmission in turbulent media \cite{Manning1986,Kon1994,Belenkii1999,Biferale2005,Robert2008,Toselli2011,Andrews2013,Cui2015,Cui2015a,Wang2017,Beason2018,Zhai2020}. For the anisotropic non-Kolmogorov turbulence model, the structure constant is a function of the anisotropic factor and the spectral power law, which reduces to the Kolmogorov structure constant when the anisotropic factor becomes 1 and the spectral power law becomes 11/3 \cite{Andrews2014}. Then the need to measure three physical quantities, namely the anisotropic non-Kolmogorov structure constant, the anisotropic factor and the spectral power law, is induced when building a model to predict anisotropic turbulence. Considering that the accurate measurement of anisotropic factor is currently a challenging task, and compared to the Kolmogorov structure constant which can be easily obtained by scintillometer, the anisotropic non-Kolmogorov structure constant is still difficult to obtain precisely, we need to find a novel method to predict anisotropic turbulence. Li et al. developed the equivalence formula between the non-Kolmogorov and the Kolmogorov structure constants, and then verified it through experimental data, which provided a method for predicting the non-Kolmogorov structure constant by utilizing the Kolmogorov structure constant \cite{Li2015}. In this paper, by establishing four different equivalence formulas between the anisotropic non-Kolmogorov and the Kolmogorov structure constants, and applying them to the extant anisotropic results, we demonstrate the possibility of using only Kolmogorov structure constant $C_n^2$ and power law $\alpha$ to predict anisotropic turbulence. And it is anticipated that this work will contribute to the comprehension of light waves transmission in the anisotropic non-Kolmogorov link and the engineering application of the anisotropic theoretical models.

\section{Formulation}

As is well-known, the power spectrum model of anisotropic non-Kolmogorov turbulence which considers the uneven distribution of the vertical and the horizontal atmosphere has the following expression \cite{Cui2015}
\begin{equation}
	{\Phi _n}\left( {\kappa ,\alpha ,\mu } \right) = A\left( \alpha  \right)C_n^2\left( {\alpha ,\mu } \right){\mu ^{2 - \alpha }}{\kappa ^{ - \alpha }},\hspace*{1em}3 < \alpha  < 4,
	\label{eq:refname1}
\end{equation}
where ${\kappa}$ is the spatial wavenumber vector, $\alpha$ is the spectral power law, $\mu$ is the anisotropic factor that is utilized to characterize the uneven distribution of turbulence cells, $C_n^2\left( {\alpha ,\mu } \right)$ is the anisotropic non-Kolmogorov structure constant that has units of $m^{3 - \alpha }$, $A\left( \alpha  \right) = \Gamma \left( {\alpha  - 1} \right)\cos \left( {\alpha \pi /2} \right)/4{\pi ^2}$, and $\Gamma \left( x \right)$ denotes the Gamma function. At $\alpha  = 11/3$ and $\mu  = 1$, Eq.~(\ref{eq:refname1}) reduce to the conventional Kolmogorov spectrum, which has the following expression
\begin{equation}
	{\Phi _n}\left( \kappa  \right) = 0.033C_n^2{\kappa ^{ - 11/3}},\hspace*{1em}1/{L_0} \ll \kappa  \ll 1/{l_0},
	\label{eq:refname2}
\end{equation}
where $C_n^2$ is the Kolmogorov structure constant that has units of $m^{ - 2/3}$, $L_0$ and $l_0$ are the outer and the inner scales. In addition, for the limiting case $C_n^2\left( {11/3,1} \right) = C_n^2$, the Kolmogorov structure constant can be obtained.

For purpose of developing the relationship between the anisotropic non-Kolmogorov and the Kolmogorov structure constants, it is necessary to start with a definition. In Baykal's paper \cite{Baykal2011}, the scintillation index of spherical wave is chosen as the basis, so we can define that the equivalent structure constants for the anisotropic non-Kolmogorov spectra are the ones that lead to the same scintillation index of spherical wave. In terms of this definition, the expression relating $C_n^2\left( {\alpha ,\mu } \right)$ to $C_n^2\left( {11/3,1} \right) = C_n^2$ is proposed.

In Ref. \cite{Andrews2014}, the scintillation index of Gaussian-beam wave for anisotropic non-Kolmogorov turbulence is obtained by Eq. (26). By setting the receiver beam parameters $\Theta  = \Lambda  = 0$ and anisotropic factors ${\mu _x} = {\mu _y} = \mu$, the scintillation index of spherical wave for anisotropic non-Kolmogorov turbulence can be expressed as
\begin{equation}
	\begin{array}{l}
		{u^2}\left( {\alpha ,\mu } \right) =  - \frac{{\rm{2}}}{\alpha }\cos \left( {\frac{{\alpha \pi }}{2}} \right)\sin \left( {\frac{{\alpha \pi }}{4}} \right){\mu ^{{\rm{2 - }}\alpha }}C_n^2\left( {\alpha ,\mu } \right){k^{3 - \frac{\alpha }{2}}}{L^{\frac{\alpha }{2}}}\\
		\hspace*{5em}\times \frac{{\Gamma \left( {\alpha  - 1} \right)\Gamma \left( {1 - \frac{\alpha }{2}} \right)\Gamma \left( {1 + \frac{\alpha }{2}} \right)\Gamma \left( {\frac{\alpha }{2}} \right)}}{{\Gamma \left( \alpha  \right)}}.
	\end{array}
	\label{eq:refname3}
\end{equation}
Here, $L$ is the propagation distance, $k = 2\pi /\lambda$, $\lambda$ is the wavelength.

For Kolmogorov turbulence, the scintillation index of spherical wave has the following expression \cite{Andrews2005}
\begin{equation}
	{u^2}\left( {11/3} \right) = 0.5C_n^2{k^{7/6}}{L^{11/6}}.
	\label{eq:refname4}
\end{equation}

By equating Eqs.~(\ref{eq:refname3}) and~(\ref{eq:refname4}), the relation between the anisotropic non-Kolmogorov structure constant $C_n^2\left( {\alpha ,\mu } \right)$ and the Kolmogorov structure constant $C_n^2$ can be obtained,
\begin{equation}
	C_n^2\left( {\alpha ,\mu } \right) = H\left( {\alpha ,\mu } \right)C_n^2,
	\label{eq:refname5}
\end{equation}
where
\begin{equation}
	H\left( {\alpha ,\mu } \right) = \frac{{ - 0.25\Gamma \left( {\alpha  + 1} \right){k^{\frac{\alpha }{2} - \frac{{11}}{6}}}{L^{\frac{{11}}{6} - \frac{\alpha }{2}}}{\mu ^{\alpha  - 2}}}}{{\Gamma \left( {\alpha  - 1} \right)\Gamma \left( {1 - \frac{\alpha }{2}} \right)\Gamma \left( {1 + \frac{\alpha }{2}} \right)\Gamma \left( {\frac{\alpha }{2}} \right)\cos \left( {\frac{{\alpha \pi }}{2}} \right)\sin \left( {\frac{{\alpha \pi }}{4}} \right)}}.
	\label{eq:refname6}
\end{equation}
In addition, when $\mu  = 1$, Eq.~(\ref{eq:refname5}) matches the non-Kolmogorov result obtained by Baykal and Gerçekcioğlu perfectly \cite{Baykal2011}.

Applying the equivalence formula~(\ref{eq:refname5}) to the generalized Rytov variance given by Andrews \cite{Andrews2014} and the anisotropic non-Kolmogorov turbulence model Eq.~(\ref{eq:refname1}) respectively, we can obtain the following formulas
\begin{equation}
	\sigma _R^2\left( \alpha  \right) = \frac{{0.5\Gamma \left( {\alpha  + 1} \right)}}{{\Gamma \left( {1 + \frac{\alpha }{2}} \right)\Gamma \left( {\frac{\alpha }{2}} \right)\alpha }}C_n^2{k^{\frac{7}{6}}}{L^{\frac{{11}}{6}}},
	\label{eq:refname7}
\end{equation}
\begin{equation}
	{\Phi _n}\left( {\kappa ,\alpha } \right) = \frac{{ - \Gamma \left( {\alpha  + 1} \right){k^{\frac{\alpha }{2} - \frac{{11}}{6}}}{L^{\frac{{11}}{6} - \frac{\alpha }{2}}}C_n^2{\kappa ^{ - \alpha }}}}{{\Gamma \left( {1 - \frac{\alpha }{2}} \right)\Gamma \left( {1 + \frac{\alpha }{2}} \right)\Gamma \left( {\frac{\alpha }{2}} \right)\sin \left( {\frac{{\alpha \pi }}{4}} \right)16{\pi ^2}}}.
	\label{eq:refname8}
\end{equation}

From Eqs.~(\ref{eq:refname7}) and~(\ref{eq:refname8}), we can find that the anisotropic factor $\mu$ has been eliminated and the anisotropic non-Kolmogorov structure constant $C_n^2\left( {\alpha ,\mu } \right)$ has been replaced by the Kolmogorov structure constant $C_n^2$, so the feasibility of using Kolmogorov structure constant $C_n^2$ and power law $\alpha$ to predict anisotropic turbulence is established. 

In Zilberman's paper \cite{Zilberman2008}, the turbulent power spectrum is chosen as the basis, and they believe that there is a specific wavenumber $\kappa _n$ to make the power spectrum for the anisotropic non-Kolmogorov and the Kolmogorov turbulence equal. Based on this assumption, we can obtain the following formula
\begin{equation}
	A\left( \alpha  \right)C_n^2\left( {\alpha ,\mu } \right){\mu ^{2 - \alpha }}\kappa _n^{ - \alpha } = A\left( {11/3} \right)C_n^2\kappa _n^{ - 11/3},
	\label{eq:refname9}
\end{equation}
where $A\left( {11/3} \right) = 0.033$. Considering that $\kappa _n$ can be chosen as ${\kappa _n} \approx {\left( {k/L} \right)^{1/2}}$ \cite{Zilberman2008}, thus the relation between the anisotropic non-Kolmogorov structure constant $C_n^2\left( {\alpha ,\mu } \right)$ and the Kolmogorov structure constant $C_n^2$ is given by
\begin{equation}
	C_n^2\left( {\alpha ,\mu } \right) = \frac{{A\left( {11/3} \right)}}{{A\left( \alpha  \right)}}{\mu ^{\alpha  - 2}}C_n^2{\left( {\sqrt {k/L} } \right)^{\alpha  - \frac{{11}}{3}}}.
	\label{eq:refname10} 
\end{equation}
In addition, when $\mu  = 1$, Eq.~(\ref{eq:refname10}) matches the non-Kolmogorov result obtained by Zilberman perfectly \cite{Zilberman2008}.

Applying the equivalence formula~(\ref{eq:refname10}) to the temporal power spectrum of irradiance fluctuations for a plane wave transmission in weak anisotropic turbulence \cite{Cui2015b} and the anisotropic non-Kolmogorov turbulence model Eq.~(\ref{eq:refname1}) respectively, we can obtain the following formulas
\begin{equation}
	\begin{array}{l}
		{W_{I\left( {pl} \right)}}\left( {\omega ,\alpha } \right) = \frac{{16{\pi ^2}A\left( {11/3} \right)C_n^2{k^{\frac{7}{6}}}{L^{\frac{{11}}{6}}}}}{{{\omega _0}}}{\left( {\frac{\omega }{{{\omega _0}}}} \right)^{1 - \alpha }}\frac{{\Gamma \left( {\frac{1}{2}} \right)\Gamma \left( {\frac{\alpha }{2} - \frac{1}{2}} \right)}}{{\Gamma \left( {\frac{\alpha }{2}} \right)}}\\
		\times {\mathop{\rm Re}\nolimits} \left\{ {1 - {}_2{F_2}\left( {1,\frac{{2 - \alpha }}{2};\frac{3}{2},\frac{{3 - \alpha }}{2};i\frac{{{\omega ^2}}}{{2\omega _0^2}}} \right) - \frac{{\Gamma \left( {\frac{\alpha }{2}} \right)\Gamma \left( {\frac{1}{2} - \frac{\alpha }{2}} \right)\Gamma \left( {\frac{1}{2} + \frac{\alpha }{2}} \right)}}{{\Gamma \left( {\frac{\alpha }{2} - \frac{1}{2}} \right)\Gamma \left( {\frac{3}{2} + \frac{\alpha }{2}} \right)}}} \right.\\
		\left. { \times {{\left( { - i\frac{{{\omega ^2}}}{{2\omega _0^2}}} \right)}^{\frac{{\alpha  - 1}}{2}}}{}_1{F_1}\left( {\frac{1}{2};\frac{{\alpha  + 3}}{2};i\frac{{{\omega ^2}}}{{2\omega _0^2}}} \right)} \right\},
	\end{array}
	\label{eq:refname11} 
\end{equation}
\begin{equation}
	{\Phi _n}\left( {\kappa ,\alpha } \right) = A\left( {11/3} \right)C_n^2{\left( {\sqrt {k/L} } \right)^{\alpha  - \frac{{11}}{3}}}{\kappa ^{ - \alpha }}.
	\label{eq:refname12} 
\end{equation}

From Eqs.~(\ref{eq:refname11}) and~(\ref{eq:refname12}), we can find that the anisotropic factor $\mu$ has been eliminated and the anisotropic non-Kolmogorov structure constant $C_n^2\left( {\alpha ,\mu } \right)$ has been replaced by the Kolmogorov structure constant $C_n^2$, so the feasibility of using Kolmogorov structure constant $C_n^2$ and power law $\alpha$ to predict anisotropic turbulence is established.

In Li's paper \cite{Li2015}, the structure function of refractive-index fluctuations is chosen as the basis, and they believe that the structure function of refractive-index fluctuations for the non-Kolmogorov and the Kolmogorov turbulence become equal when the separation distance $r = {L_0}$. However, the anisotropy and the variation of power law will change the distribution of refractive-index in the atmosphere, which in turn affects the value of structure function. Therefore, we choose the turbulent power spectrum instead of the structure function of refractive-index fluctuations as the basis and define that when the separation distance $r = {L_0}$, the power spectrum for the anisotropic non-Kolmogorov and the Kolmogorov turbulence become equal. Then the following formula can be obtained,
\begin{equation}
	A\left( \alpha  \right)C_n^2\left( {\alpha ,\mu } \right){\mu ^{2 - \alpha }}L_0^\alpha  = A\left( {11/3} \right)C_n^2L_0^{11/3},
	\label{eq:refname13} 
\end{equation}
where the spatial wavenumber vector $\kappa  = 1/r$. Accordingly, the relation between the anisotropic non-Kolmogorov structure constant $C_n^2\left( {\alpha ,\mu } \right)$ and the Kolmogorov structure constant $C_n^2$ can be expressed as
\begin{equation}
	C_n^2\left( {\alpha ,\mu } \right) = \frac{{A\left( {11/3} \right)}}{{A\left( \alpha  \right)}}C_n^2{\mu ^{\alpha  - 2}}L_0^{11/3 - \alpha }.
	\label{eq:refname14} 
\end{equation}

Applying the equivalence formula~(\ref{eq:refname14}) to the temporal power spectrum of irradiance fluctuations for a plane wave transmission in weak anisotropic turbulence \cite{Cui2015b} and the anisotropic non-Kolmogorov turbulence model Eq.~(\ref{eq:refname1}) respectively, we can obtain the following formulas
\begin{equation}
	\begin{array}{l}
		{W_{I\left( {pl} \right)}}\left( {\omega ,\alpha } \right) = \frac{{16{\pi ^2}A\left( {11/3} \right)C_n^2{k^{3 - \frac{\alpha }{2}}}{L^{\frac{\alpha }{2}}}L_0^{11/3 - \alpha }}}{{{\omega _0}}}{\left( {\frac{\omega }{{{\omega _0}}}} \right)^{1 - \alpha }}\frac{{\Gamma \left( {\frac{1}{2}} \right)\Gamma \left( {\frac{\alpha }{2} - \frac{1}{2}} \right)}}{{\Gamma \left( {\frac{\alpha }{2}} \right)}}\\
		\times {\mathop{\rm Re}\nolimits} \left\{ {1 - {}_2{F_2}\left( {1,\frac{{2 - \alpha }}{2};\frac{3}{2},\frac{{3 - \alpha }}{2};i\frac{{{\omega ^2}}}{{2\omega _0^2}}} \right) - \frac{{\Gamma \left( {\frac{\alpha }{2}} \right)\Gamma \left( {\frac{1}{2} - \frac{\alpha }{2}} \right)\Gamma \left( {\frac{1}{2} + \frac{\alpha }{2}} \right)}}{{\Gamma \left( {\frac{\alpha }{2} - \frac{1}{2}} \right)\Gamma \left( {\frac{3}{2} + \frac{\alpha }{2}} \right)}}} \right.\\
		\left. { \times {{\left( { - i\frac{{{\omega ^2}}}{{2\omega _0^2}}} \right)}^{\frac{{\alpha  - 1}}{2}}}{}_1{F_1}\left( {\frac{1}{2};\frac{{\alpha  + 3}}{2};i\frac{{{\omega ^2}}}{{2\omega _0^2}}} \right)} \right\},
	\end{array}
	\label{eq:refname15} 
\end{equation}
\begin{equation}
	{\Phi _n}\left( {\kappa ,\alpha} \right) = A\left( {11/3} \right)C_n^2L_0^{11/3 - \alpha }{\kappa ^{ - \alpha }}.
	\label{eq:refname16} 
\end{equation}

From Eqs.~(\ref{eq:refname15}) and~(\ref{eq:refname16}), we can find that the anisotropic factor $\mu$ has been eliminated and the anisotropic non-Kolmogorov structure constant $C_n^2\left( {\alpha ,\mu } \right)$ has been replaced by the Kolmogorov structure constant $C_n^2$, so the feasibility of using Kolmogorov structure constant $C_n^2$ and power law $\alpha$ to predict anisotropic turbulence is established.

In order to better describe the anisotropic turbulence, Andrews et al. proposed a new power spectrum model, which has two anisotropic factors to parameterize the asymmetry of turbulence eddies in the orthogonal $xy$-plane. And it has the following expression \cite{Andrews2014}
\begin{equation}
	{\Phi _n}\left( {\kappa ,\alpha ,{\mu _x},{\mu _y}} \right) = \frac{{A\left( \alpha  \right)C_n^2\left( {\alpha ,{\mu _x},{\mu _y}} \right){\mu _x}{\mu _y}}}{{{{\left( {\mu _x^2\kappa _x^2 + \mu _y^2\kappa _y^2} \right)}^{\alpha /2}}}},3 < \alpha  < 4,
	\label{eq:refname17} 
\end{equation}
where $\kappa _x$ and $\kappa _y$ are the $x$ and $y$ components of the spatial wavenumber vector $\kappa$, $C_n^2\left( {\alpha ,{\mu _x},{\mu _y}} \right)$ is the anisotropic non-Kolmogorov structure constant that has units of $m^{3 - \alpha }$, $\mu _x$ and $\mu _y$ are the anisotropic factors. If we assume that symmetric distribution of turbulence cells in the plane orthogonal to the direction of propagation $z$, i.e., the particular case ${\mu _x} = {\mu _y} = \mu$, the spectrum model can be reduced to Eq.~(\ref{eq:refname1}).

In Ref. \cite{Andrews2014}, the scintillation index of Gaussian-beam wave for anisotropic non-Kolmogorov turbulence is obtained by Eq. (26). By setting the receiver beam parameters $\Theta  = \Lambda  = 0$, the scintillation index of spherical wave for anisotropic non-Kolmogorov turbulence can be expressed as
\begin{equation}
	\begin{array}{l}
		{u^2}\left( {\alpha ,{\mu _x},{\mu _y}} \right) =  - \frac{1}{{\alpha \pi }}\cos \left( {\frac{{\alpha \pi }}{2}} \right)\sin \left( {\frac{{\alpha \pi }}{4}} \right)C_n^2\left( {\alpha ,{\mu _x},{\mu _y}} \right){k^{3 - \frac{\alpha }{2}}}{L^{\frac{\alpha }{2}}}\\
		\times \frac{{\Gamma \left( {\alpha  - 1} \right)\Gamma \left( {1 - \frac{\alpha }{2}} \right)\Gamma \left( {1 + \frac{\alpha }{2}} \right)\Gamma \left( {\frac{\alpha }{2}} \right)}}{{\Gamma \left( \alpha  \right)}}\int_0^{2\pi } {{{\left( {\frac{{{{\cos }^2}\theta }}{{\mu _x^2}} + \frac{{{{\sin }^2}\theta }}{{\mu _y^2}}} \right)}^{\frac{\alpha }{2} - 1}}d\theta }.
	\end{array}
	\label{eq:refname18} 
\end{equation}

According to the previous idea, by choosing the scintillation index of spherical wave as the basis and equating Eqs.~(\ref{eq:refname18}) and~(\ref{eq:refname4}), we can obtain the relation between the anisotropic non-Kolmogorov structure constant $C_n^2\left( {\alpha ,{\mu _x},{\mu _y}} \right)$ and the Kolmogorov structure constant $C_n^2$,
\begin{equation}
	C_n^2\left( {\alpha ,{\mu _x},{\mu _y}} \right) = \frac{{M\left( \alpha  \right)}}{{G\left( {\alpha ,{\mu _x},{\mu _y}} \right)}}C_n^2,
	\label{eq:refname19} 
\end{equation}
where
\begin{equation}
	M\left( \alpha  \right) = 0.5\Gamma \left( {\alpha  + 1} \right)\pi {k^{\frac{\alpha }{2} - \frac{{11}}{6}}}{L^{\frac{{11}}{6} - \frac{\alpha }{2}}},
	\label{eq:refname20} 
\end{equation}
\begin{equation}
	\begin{array}{l}
		G\left( {\alpha ,{\mu _x},{\mu _y}} \right) =  - \Gamma \left( {\alpha  - 1} \right)\Gamma \left( {1 - \frac{\alpha }{2}} \right)\Gamma \left( {1 + \frac{\alpha }{2}} \right)\Gamma \left( {\frac{\alpha }{2}} \right)\\
		\times \cos \left( {\frac{{\alpha \pi }}{2}} \right)\sin \left( {\frac{{\alpha \pi }}{4}} \right)\int_0^{2\pi } {{{\left( {\frac{{{{\cos }^2}\theta }}{{\mu _x^2}} + \frac{{{{\sin }^2}\theta }}{{\mu _y^2}}} \right)}^{\frac{\alpha }{2} - 1}}d\theta }.
	\end{array}
	\label{eq:refname21} 
\end{equation}
In addition, when ${\mu _x} = {\mu _y} = \mu$, Eq.~(\ref{eq:refname19}) can be reduced to Eq.~(\ref{eq:refname5}).

Applying the equivalence formula~(\ref{eq:refname19}) to the generalized Rytov variance given by Andrews \cite{Andrews2014} and the temporal power spectrum of AOA fluctuations for a plane wave transmission in weak anisotropic turbulence by Cui \cite{Cui2018} respectively, we can obtain the following formulas
\begin{equation}
	\sigma _R^2\left( {\alpha} \right) = \frac{{0.5\Gamma \left( {\alpha  + 1} \right)}}{{\Gamma \left( {1 + \frac{\alpha }{2}} \right)\Gamma \left( {\frac{\alpha }{2}} \right)\alpha }}C_n^2{k^{\frac{7}{6}}}{L^{\frac{{11}}{6}}},
	\label{eq:refname22} 
\end{equation}
\begin{equation}
	\begin{array}{l}
		{W_{\theta \left( {pl} \right)}}\left( {\omega ,\alpha} \right) = \frac{{0.25\Gamma \left( {\alpha  + 1} \right){k^{\frac{\alpha }{4} - \frac{{13}}{{12}}}}{L^{\frac{{25}}{{12}} - \frac{\alpha }{4}}}}}{{ - \Gamma \left( {1 - \frac{\alpha }{2}} \right)\Gamma \left( {1 + \frac{\alpha }{2}} \right)\Gamma \left( {\frac{\alpha }{2}} \right)\sin \left( {\frac{{\alpha \pi }}{4}} \right)}}C_n^2\\
		\frac{1}{{{\omega _0}}}\Gamma \left( {\frac{1}{2}} \right){\left( {\frac{\omega }{{{\omega _0}}}} \right)^{\frac{{1 - \alpha }}{2}}}{\left( {\frac{{{c^2}{D^2}}}{4}} \right)^{\frac{{\alpha  - 5}}{4}}}\exp \left( { - \frac{{{c^2}{D^2}k{\omega ^2}}}{{8L\omega _0^2}}} \right)\\
		\times \left\{ {\left[ {1 - \cos \left( {2\beta } \right)} \right]{W_{\frac{{3 - \alpha }}{4},\frac{{3 - \alpha }}{4}}}\left[ {\left( {\frac{{{c^2}{D^2}}}{4}} \right)\frac{{k{\omega ^2}}}{{L\omega _0^2}}} \right] + 2\frac{\omega }{{{\omega _0}}}\cos \left( {2\beta } \right){{\left( {\frac{k}{L}} \right)}^{\frac{1}{2}}}} \right.\\
		\left. { \times {{\left( {\frac{{{c^2}{D^2}}}{4}} \right)}^{\frac{1}{2}}}{W_{\frac{{1 - \alpha }}{4},\frac{{1 - \alpha }}{4}}}\left[ {\left( {\frac{{{c^2}{D^2}}}{4}} \right)\frac{{k{\omega ^2}}}{{L\omega _0^2}}} \right]} \right\}.
	\end{array}
	\label{eq:refname23} 
\end{equation}

From Eqs.~(\ref{eq:refname22}) and~(\ref{eq:refname23}), we can find that the anisotropic factors $\mu _x$ and $\mu _y$ have been eliminated and the anisotropic non-Kolmogorov structure constant $C_n^2\left( {\alpha ,{\mu _x},{\mu _y}} \right)$ has been replaced by the Kolmogorov structure constant $C_n^2$, so the feasibility of using Kolmogorov structure constant $C_n^2$ and power law $\alpha$ to predict anisotropic turbulence is established.

In fact, Beason also gives the relationship between the anisotropic non-Kolmogorov structure constant $C_n^2\left( {\alpha ,{\mu _x},{\mu _y}} \right)$ and the Kolmogorov structure constant $C_n^2$ \cite{Beason2019}. Then applying it to the generalized Rytov variance and the temporal power spectrum of AOA fluctuations for plane wave, we can also find the elimination phenomenon, so our conclusion is confirmed. In addition, although the form differs greatly, the calculation result of Eq.~(\ref{eq:refname19}) is consistent with Beason's paper.

\section{Conclusion}

Based on three modeling approaches, four different equivalence formulas between the anisotropic non-Kolmogorov and the Kolmogorov structure constants are derived in this paper. And applying them to the extant anisotropic results, the anisotropic factor is eliminated and the possibility of using only Kolmogorov structure constant $C_n^2$ and power law $\alpha$ to predict anisotropic turbulence is established. It should be noted that this work will contribute to the comprehension of light waves transmission in the anisotropic non-Kolmogorov link and the engineering application of the anisotropic theoretical models.

\section{Disclosures}

The authors declare no conflicts of interest.

\section{Acknowledgments}

This work was supported by the National Natural Science Foundation of China under Grant 61901097.

\bibliography{RPLib}

\end{document}